# Diabatic quantum gates

Pedro J. Salas Peralta

Departamento de Tecnologías Especiales Aplicadas a la Telecomunicación, Universidad Politécnica de Madrid, Ciudad Universitaria s/n, 28040 Madrid (Spain)

*email: psalas@etsit.upm.es*

**Abstract**
At present, several models for quantum computation have been proposed. Adiabatic quantum computation scheme particularly offers this possibility and is based on a slow enough time evolution of the system, where no transitions take place. In this work, a new strategy for quantum computation is provided from the opposite point of view. The objective is to control the non-adiabatic transitions between some states in order to produce the desired exit states after the evolution. The model is introduced by means of an analogy between the adiabatic quantum computation and an inelastic atomic collision. By means of a simple two-state model, several quantum gates are reproduced, concluding the possibility of diabatic universal fault-tolerant quantum computation. Going a step further, a new quantum diabatic computation model is glimpsed, where a carefully chosen Hamiltonian could carry out a non-adiabatic transition between the initial and the sought final state.

## 1. Introduction

The adiabatic theorem has a long history in quantum mechanics [1]. It was Farhi *et al* [2] who applied it to quantum computation by suggesting a new way of solving classic problems, such as SAT, by means of an algorithm based on adiabatic quantum computation (AQC). After the first speculation about the promising power of this novel way of computation [3], finally demonstrated its polynomial equivalence with the conventional circuit model [4]. In spite of that, it has some advantages such as an inherent robustness to some kinds of noise. AQC is robust against dephasing in the ground state or thermal errors [5, 6]. Several error correction strategies have also been proposed [7]. From the experimental point of view, some quantum adiabatic algorithms have already been implemented experimentally by means of NMR techniques [8].

In the following, the adiabatic quantum computation model is briefly described, and the details may be reviewed elsewhere [9]. Consider a state evolving according the Schrödinger equation and a time-depending Hamiltonian H(t) (t ∈ [0,T], T being the total computation time), having a set of eigenstates and eigenvalues $\{|\psi_k(t)>, E_k(t)\}$. The adiabatic theorem states that if the initial state is the kth-eigenvector $|\psi_k(t=0)>$ and H(t) varies slowly enough, the instantaneous state $|\psi(0 \leq t \leq T)>$ of the system will remain close to the state $|\psi_k(t=T)>$ at the end of the process. Starting in the ground state $|\psi_0(t=0)>$, the adiabatic evolution will guarantee that the system will be (with a great probability) in the same state at the end. More precisely, defining the minimum gap between the lowest two energy eigenvalues as:

$$\Delta E_{min} = \min_{0 \leq t \leq T} (E_1(t) - E_0(t)) \tag{1}$$

and the maximum value of the operator dH(t)/dt between these two states

$$D_{max} = \max_{0 \leq t \leq T} \left| \langle \psi_0(t) | \frac{dH}{dt} | \psi_1(t) \rangle \right|, \tag{2}$$

if the initial state is $|\psi_0(0)>$, provided the condition

$$\frac{D_{max}}{\Delta E_{min}^2} \leq \varepsilon \tag{3}$$

is fulfilled, then

$$\left| \langle \psi_0(T) | \psi(T) \rangle \right|^2 \geq 1 - \varepsilon^2 \tag{4}$$

If H(T) = H(0), then the final state $|\psi(T)>$ is $\varepsilon^2$-close (according (4)) to the $|\psi_0(T)>$ state, except for an overall phase. Roughly speaking, the meaning of this statement is: the state of the system will be kept close to the instantaneous state $|\psi_0(t)>$ if it is weakly coupled with the remaining states (small $D_{max}$) and is kept well separated in energy from them (large $\Delta E_{min}$). In this situation, there are neither crossings nor avoided-crossings between the considered states and, consequently, the transition probabilities to the excited states are very small.

Despite the above criterion not actually being necessary nor sufficient in general, it has been widely used because of its simplicity. Work has recently been carried out in order to study the consistency of the theorem [10], and replace the previous formulation with a rigorous statement [11], even including noise [12].

The adiabatic theorem can be used to design a new paradigm of quantum computation. The model is specified by two Hamiltonians (in the simplest case), the initial $H_0$ and final $H_1$. The initial state is the ground state of $H_0$ that is required to be an-easy-to-prepare state whereas the solution of the computation is the ground state of $H_1$. The problem could be to work out the structure of $H_0$ and $H_1$. The time evolution is controlled by means of the total Hamiltonian (H) prepared as the interpolation of the previous Hamiltonians, depending on a parameter s(t): $H(t) = f(s(t)) H_0 + g(s(t)) H_1$, with $s \in [0, 1]$ and $s = t/T$, T being the total computation time. The local condition is usually necessary for the Hamiltonian (requiring that its implementation only involve a constant number of particles), in order to be realistically implemented.

Initially, the system is synthesized in the ground state of $H_0$, $|\psi_0(0)>$, and then evolves according to H(t). If the evolution rate (ds(t)/dt) is slow enough (adiabatic evolution), the intermediate state ($|\psi(t)>$) will not produce any transition to (possible) higher energy states, and will end in the ground state of $H_1$, that has been chosen as the solution of the problem.

The question addressed in this work is: could the main framework of AQC be used to reach the solution of a problem by taking advantage of the possible transitions between the states? In fact, this is what is happening in the well-known context of quantum chemistry when two atoms come close to form a molecule, if the Born-Oppenheimer approximation [13] is not fulfilled. This behaviour is crucial in the context of inelastic atomic collisions [14], where the studied processes are, specifically, those producing outgoing states different from the ingoing ones. The transitions are produced by the breakdown of the Born-Oppenheimer approximation, involving states whose energies cross or pseudocrosse (avoided crossing). The molecular Hamiltonian depends on the atomic separation R(t) that can be seen as a time-dependent parameter of the total Hamiltonian. The starting point for studying these systems is to find the instantaneous (for R constant) adiabatic states and then solving the time-independent Schrödinger equation. The dynamic is included by expanding the total collision state in the adiabatic basis set. If, through the evolution, a transition probability from the initial state to some of the final states is not negligible, it is said that a non-adiabatic transition may have occurred. In this work, the strongly non-adiabatic behaviour of the states is proposed as a possible mechanism for quantum computation, calling it diabatic quantum computation (as opposed to adiabatic); a term used in atomic collisions, although not exactly with the same meaning. In this context, the word diabatic refers to a new set of basis states providing smaller coupling values. This new basis includes states that go smoothly across an avoided-crossing and are more suitable to describe inelastic processes at medium and higher energies.

As a result of the above ideas on non-adiabatic transitions (diabatic behaviour) between quantum states, the construction of quantum diabatic gates will be established. The main framework of AQC can be used but with some new characteristics. In the AQC, by choosing the ending Hamiltonian properly, the final correct state is identified with the ground state. Now, in the diabatic behaviour, the correct state will be the first excited one of a suitable Hamiltonian. In order to get this diabatic evolution, the speed of the computation is not restricted to being a sufficiently small evolution speed as in AQC. In contrast, the final state will be reached with high probability if the computation speed is high enough. This diabatic behaviour will not limit the computation time as in AQC. Another advantage of the diabatic computation is its robustness against errors shown as an intrinsic fault-tolerance, as established in section 3.1. In addition, the concept of diabatic computation as a sequence of diabatic gates is suitable to be extended to a full diabatic computation, understood as a single evolution carried out by means of an appropriate Hamiltonian. The only remaining work is to characterise the physical processes by implementing this evolution. These Hamiltonians could involve non-local interactions of more than two or three qubits. This could be seen as a heavy restriction nowadays, but is not forbidden by the laws of Quantum Mechanics. Providing suitable quantum systems as well as general non-local Hamiltonians is not the scope of the paper.

The paper is structured as follows. In Sec. 2 the main ideas about inelastic atomic collisions are reviewed and are related to the diabatic computation model. In Sec. 3, a simple two-state model will be used to implement several diabatic quantum gates, demonstrating its general intrinsic fault-tolerance and estimating the error gate probability for each one.

## 2. Diabatic quantum computation model

The adiabatic computation involves the slow enough evolution of a time-dependent Hamiltonian, this fact ensures the lack of transitions. The opposite behaviour is sought in the present method. The goal is to develop the model through a parallelism with an inelastic atomic collision.

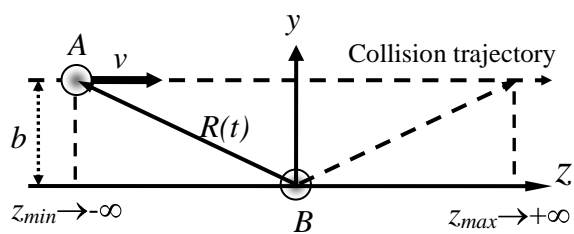

**Fig. 1** Reference framework to describe an atomic collision between two atoms A and B. In the process, the ingoing atom A goes from z → -∞ to z → +∞ through a straight line trajectory.

Consider an inelastic atomic collision where an atom A crosses an interaction region occupied by another atom B or a field. The collision frame is shown in figure 1. The incoming atom comes from $z_{min}$ → -∞ (equivalently t → -∞, because the time origin is situated on the coordinate z=0) to $z_{max}$ → +∞ (or t → +∞). If the atomic speed (v) is large enough, the trajectories can be taken as straight lines and characterized by an impact parameter b. These are (roughly) the assumptions made in the Impact Parameter Method [15], used to describe atomic collisions in the medium energy range. Notice that the parameter R(t) provides a radial speed as $\dot{R}(t)$ =zv/R, with z=vt fulfilling the equation $R(t)^2 = z(t)^2 + b^2$.

The total Hamiltonian H describing the collision, as well as the states are time-dependent through the implicit R(t) parameter dependence. To describe the dynamic, the collision state |ψ(t)> is developed in a basis set {|$μ_k$(R(t))>}:

$$|\psi(t)\rangle = \sum_k a_k(t)|\mu_k(R(t))\rangle \qquad (5)$$

It is evident that in the case of a "real" atomic collision, all of the states must depend on the electronic coordinates; however these are not shown in the equations in order to introduce a closer notation to that used below.

The Hamiltonian H is not (in general) diagonal in the basis set {|$μ_k$>}, then, in the Molecular Model of atomic collisions [14, 15], a new and more appropriate adiabatic (orthonormal) basis set {|$φ_k$(R)>} is provided as the instantaneous eigenvectors of H. This process is static since the diagonalizing process (H|$φ_k$(R)> = $E_k$(R)|$φ_k$(R)>) is carried out for each constant R value. The dynamic is included by solving the standard time-dependent Schrödinger equation for the collision state of the system |ψ(t)> expanded in the adiabatic basis set. Introducing |ψ(t)> into the Schrödinger equation, the following (general) differential coupled equation system is obtained:

$$\frac{da_k(t)}{dt} = -\sum_n a_n(t)\,[\langle\phi_k|\partial/\partial t|\phi_n\rangle + i\langle\phi_k|H|\phi_n\rangle] \qquad (6)$$

The system involves the dynamic coupling <$φ_k$|∂/∂t|$φ_n$> and the electrostatic coupling <$φ_k$|H|$φ_n$>. The index k runs over the coupled states. Two properties simplify the aforementioned equations: using an adiabatic

basis set, $<\phi_k|H|\phi_n> = E_n \delta_{kn}$, and the matrix elements $<\phi_k|\partial/\partial t|\phi_k> = [<\partial\phi_k/\partial t|\phi_k>+<\phi_k|\partial/\partial t|\phi_k>]/2 = (1/2) \partial<\phi_k|\phi_k>/\partial t = 0$ (bearing in mind that the states $|\phi_k>$ are real functions). Taking into account the relationship $z = vt$, the above system of equations can be transformed into another equivalent depending on the z coordinate. In this context, the Massey parameter [16] is interesting in classifying the processes according the transition probabilities. It comes from very general considerations. The non-adiabatic behaviour (large transition probability) between two states is expected to be important when the typical collision frequency written as the inverse of the collision time ($\omega = 1/\tau c$), matches the frequency of the energy splitting ($\Delta E/\hbar$). Taking into account that $\tau c \sim a/v$ (a being a characteristic z-region of interaction), the constraint $\xi = a\Delta E/v \sim 1$ (with $\hbar = 1$) is obtained. The condition of adiabaticity comes from a small coupling (low transition probability), implying a large value of $\Delta E$ and small v, then $\xi >> 1$. The speed for the maximum transition probability can be estimated be means of this Massey parameter $\xi$. The previous behaviour is closely related to the coupled system (6), if the following change in the variable $z = vt$ is introduced, and an energy phase is extracted:

$$a_k(z) = b_k(z) \exp\left(-\frac{i}{v} \int_{-\infty}^{z} E_k(R(z')) dz'\right) \qquad (7)$$

The new system of equations is:

$$\frac{db_k(z)}{dz} = -\sum_{n \neq k} b_n(z) \frac{1}{v} \langle\phi_k|\partial/\partial t|\phi_n\rangle \exp\left(-\frac{i}{v} \int_{-\infty}^{z} [(E_n(R(z')) - E_k(R(z')) dz']\right)$$

(8)

The coupling $<\phi_k|\partial/\partial t|\phi_n>$ should be expressed in terms of z, with two new couplings appearing: the radial coupling $<\phi_k|\partial/\partial R|\phi_n>$ and a rotational one. The radial coupling originates in the case of dealing with states of the same molecular symmetry and appears, for example, in an avoided-crossing. The rotational coupling comes out between the states of different molecular symmetry and will not be taken into account in the following, because the states considered will have the same symmetry and display an avoided-crossing.

The system of equations has, in general, to be solved numerically. The behaviour of the adiabaticity follows the tracks of the Massey parameter. If the initial conditions are $b_k(0) = \delta_{k0}$, a simple first order estimation for the transition probability will depend on how quickly the phase oscillates. An estimation of this phase could be done through the $\xi$ parameter. The coefficient $b_{m \neq 0}(+\infty)$ will be small if $\xi \gg 1$ [17] and the behaviour is adiabatic. In the case of $\xi \sim 1$, a large value of $b_{m \neq 0}(+\infty)$ is expected, and the behaviour is non-adiabatic.

In principle, the most appropriate choice as a basis set to develop the collision state is the adiabatic one. To describe an inelastic collision adequately, all coupled states must be included in order to account for the possible transitions. Unfortunately, in some cases, when the states are highly coupled or when the speed of the colliding systems is high enough, the couplings are very active to produce transitions. In this case the number of states to be included in the system (8) is so huge that it makes it impossible to carry out the description. In this situation, a new (and smaller) set of states, called diabatic, could help to decrease the dimension of the problem. Diabatic states were introduced in the early 60's by Lichten [18] and generalized by Smith [19]. Roughly speaking, they have the property of running smoothly through an avoided-crossing, having smaller couplings. Intensive work has been carried out in the past on the adequate definition of diabatic states (see [14] and references therein).

Trying to introduce the flavour of an atomic collision in the present computation model, a parameter $s(t)$ similar to $R(t)$, is included, allowing the computation to be seen as a kind of "collision". Defining T as the total computation time, the parameter s is defined as $s(t) = t/T$. The states for the incoming particle (for R large in the atomic collision model), would now correspond to $s = 1$ and the closest atomic distance (equivalent to R small), to $s = 0$. Suppose the searched for state is one of the eigenstates $|\chi_m\rangle$ of a non-degenerate Hamiltonian $H_0$. The diabatic quantum computation would be like a kind of "collision" depending on the $s(t)$ parameter. By keeping the same framework as in the AQC scheme, two time-independent Hamiltonians are taken into account: $H_0$ (carrying out the information of the initial and/or final possible states) and $H_W$ (including a coupling). The $H_W$ includes a coupling between the basis eigenstates of $H_0$, allowing the possibility of transitions between them. The evolution of the system is produced according the Hamiltonian:

$$H(s(t)) = f(s(t))\, H_0 + g(s(t))\, H_W \qquad (9)$$

The functions $f(s)$ and $g(s)$ are chosen to fulfil the general properties:

$$0 \xleftarrow{s \to 0} |f(s)| \xrightarrow{s \to 1} 1$$
$$1 \xleftarrow{s \to 0} |g(s)| \xrightarrow{s \to 1} 0 \qquad (10)$$

The behaviour reflects the introduction of the perturbation $H_W$, when s goes from 1 to 0. The state vector of this "collision" is expanded in the adiabatic basis states (eigenvectors) $\{|\phi_k(s)\rangle\}$ of $H(s)$:

$$\left|\psi(t)\right\rangle = \sum_k a_k(t)\left|\phi_k(s(t))\right\rangle \qquad (11)$$

The time-evolution is controlled by a system of coupled equations similar to (6) or (8) if the z variable is taken. The diabatic computation is seen as a process in which the synthesis of the appropriate ingoing state (expanded in the adiabatic basis of $H_0$), is introduced for the s(0) = 1 black box wire, and then the "collision" run evolving forward and backward from s = 1 → 0 → 1, and get the outgoing state $|\phi_{out}(s\to1)>$. For s → 1, the coupling goes to 0 because g(s→1) → 0, then $|\phi_{out}(s\to1)>$ could be very close to the sought eigenvector $|\chi_m>$ of $H_0$ (or perhaps some linear combination of them), if values of the parameters (v, b) are carefully chosen.

2.1 Two-state model

In general, the diabatic quantum computation could involve ingoing and outgoing states developed in the adiabatic basis set $\{|\phi_k>, k=1,..,n > 2\}$, but, with an adequate choice of the Hamiltonian and the parameters (v, b), only a two-state model could be enough. This would be case if the two coupled states considered are far separated in energy from the remaining excited states or if they are not coupled by symmetry considerations. In this case the process of computation involves only two eigenstates $\{|\chi_0>, |\chi_1>\}$ of the Hamiltonian $H_0$:

$$H_0\left|\chi_0\right\rangle = \epsilon_0\left|\chi_0\right\rangle \quad H_0\left|\chi_1\right\rangle = \epsilon_1\left|\chi_1\right\rangle \qquad (12)$$

with $H_W$ providing the coupling between them. These states are not (in general) eigenstates of the total Hamiltonian H = f(s) $H_0$ + g(s) $H_W$, so the energies $\epsilon_1$ and $\epsilon_2$ could cross. Diagonalizing H in the basis $\{|\chi_0>, |\chi_1>\}$ a new adiabatic basis set $\{|\phi_0>, |\phi_1>\}$ is obtained whose new eigenvalues ($E_{0,1}(s)$) could show an avoided crossing (pseudocrossing). The adiabatic basis is used to develop the evolution of the total state vector $|\psi(t)>$:

$$\left|\psi(t)\right\rangle = a_0(t)\left|\phi_0(s(t))\right\rangle + a_1(t)\left|\phi_1(s(t))\right\rangle \qquad (13)$$

The coefficients $\{a_0(t), a_1(t)\}$ are provided by the general equation system (6). Changing to the z variable through z = vt, the two-state coupled equation system is:

$$\frac{da_0(z)}{dz} = -i\, a_0(z)\, \frac{E_0(s(z))}{v} - a_1(z)\, \frac{z}{s(z)}\, W(s(z))$$

$$\frac{da_1(z)}{dz} = -i\, a_1(z)\, \frac{E_1(s(z))}{v} + a_0(z)\, \frac{z}{s(z)}\, W(s(z)) \tag{14}$$

$W(s(z)) = <\phi_0(s)|\partial/\partial s|\phi_1(s)>$ being a kind of "radial coupling" in the parameter s (having considered the property $<\phi_0(s)|\partial/\partial s|\phi_1(s)> = -<\phi_1(s)|\partial/\partial s|\phi_0(s)>$). The first terms on the right hand side of (14) describe the adiabatic evolution (keeping the populations and, perhaps, changing the phases) and, the second terms include the non-adiabatic transitions by means of the W coupling. The integration of this system is carried out according to take s = 1: from $z_{min} = -(1-b^2)^{1/2}$ to $z_{max} = (1-b^2)^{1/2}$, b being the impact parameter, seen now as a parameter to be adjusted.

### 3. Diabatic gates

The first step in reaching quantum computation is to describe the quantum diabatic gates. Each gate can be seen as a black box inside which a quantum evolution, according to a Hamiltonian, takes place in some ingoing state in order to produce an outgoing one. Unlike the adiabatic computation, some transitions are required and, to reach them, two parameters (v, b) are free to be adjusted.

Working in the computation basis set $\{|\chi_0> = |0>, |\chi_1> = |1>\}$, the initial (s = 1) state is introduced into the black-box-gate. The whole evolution as a "collision" varying s = 1 → 0 → 1, is described through an evolution operator U(t,-t) (remember the time origin t = 0 is in z = 0) from t = -T to t = T in the adiabatic basis set [20]:

$$U(T,-T) = \begin{pmatrix} (1-p)\, e^{i2\alpha_{00}} + p e^{i2\alpha_{01}} & -2i\sqrt{(1-p)p}\, \sin(\alpha_{00} - \alpha_{01}) \\ -2i\sqrt{(1-p)p}\, \sin(\alpha_{00} - \alpha_{01}) & (1-p)\, e^{-i2\alpha_{00}} + p e^{-i2\alpha_{01}} \end{pmatrix} \tag{15}$$

p being the probability of a non-adiabatic transition through the avoided-crossing and $\alpha_{ij}$ some phases, both depending on the (v, b) values. The probability that the system exits through a particular outgoing channel can be calculated by applying the evolution operator U(T,-T) to the incoming state. The different quantum diabatic gates can be represented by choosing the values of the parameters (v, b) adequately as will be shown throughout paragraphs 3.2-3.5. First of all the fault tolerance of the diabatic gates (15) will be considered.

## 3.1 Fault-tolerance of diabatic gates

The features providing the power to quantum computers are parallelism and interference, which are intrinsically quantum properties. The implementation of a quantum algorithm requires a quantum computer to keep working on the qubits for a long time. The computation requires the creation and manipulation of entangled states involving large ensembles of qubits. Unfortunately the quantum states are necessarily coupled with the environment producing the qubit decoherence. This process introduces errors into the computation, making it useless. To fight against error accumulation, Shor [21, 22] and Steane [23] introduced in the mid 90's, the concept of quantum error correcting codes, capable of keeping the quantum decoherence under control. Unfortunately, error-correcting methods are not strong enough to achieve a total control of error spreading through a quantum algorithm. In trying to solve this problem, Shor introduced fault-tolerant methods [24] in quantum computation. There are several basic ideas involved in it: applying quantum gates directly to the encoded qubits, correcting the errors periodically, a carefully designed encoded gates in order to avoid the error spreading and the use of a concatenated quantum code structure with a hierarchical encoding [25]. Roughly speaking, a fault-tolerant recovery method would introduce fewer errors than those it is able to eliminate. The fusion of fault-tolerant encoded quantum gates and concatenated codes has established the existence of an error threshold. If the evolution and gate errors are below this threshold, quantum states will remain stabilized for a time long enough to carry out the computation. Several estimations for the value of this threshold have been published [26, 27, 28, 29, 30, 31]. From these works, it is possible to establish an approximate error-gate-probability-threshold ($P_{err}$) to carry out a long enough quantum computation, as $P_{err} \leq 10^{-4}$.

Following the same argument used in [32], the gate error probability is defined in the following way: consider the target unitary operation to be implemented as represented by the evolution operator $U_t$ and the approximate one by $U_a$. The error included on $U_a$ could come from an error in the parameters (v, b) selected experimentally. Calling |Ψ> the initial vector state, the gate error probability ($P_e(\Psi)$) associated is defined as the orthogonal component of the vector $\left|\xi_\psi\right\rangle = (U_a - U_t) |\Psi\rangle = D |\Psi\rangle$:

$$P_e(\Psi) = \left\langle \xi_\Psi^\perp | \xi_\Psi^\perp \right\rangle \qquad (16)$$

The error probability for the gate is defined as:

$$P_e = \max_{\forall |\Psi\rangle} (P_e(\Psi)) \qquad (17)$$

and fulfils the condition:

$$P_e(\Psi) = \langle \xi_\Psi^\perp | \xi_\Psi^\perp \rangle \leq \langle \xi_\Psi | \xi_\Psi \rangle = tr(\rho_\Psi P) \qquad (18)$$

P being the positive and hermitian operator $D^+D$ and $\rho_\Psi = |\Psi\rangle\langle\Psi|$. By diagonalizing P and taking into account that $tr\{\rho_\Psi P\}$ is upper bounded by $d_m$ = max$\{d_i$, eigenvalues of P$\}$ for every state vector $|\Psi\rangle$, the error probability for the gate is $P_e \leq d_m$. As the P operator is positive, the condition $d_m \leq tr\{P\}$ is also fulfilled, then:

$$P_e \leq d_m \leq tr(P) \qquad (19)$$

In some cases the tr$\{P\}$ is much easier to be calculated than $d_m$ and it will be used as an upper bound of $P_e$.

The next step will be to check the error propagation when the gate is described by the evolution operator U(T,-T). As was mentioned before, the error affecting the operator $U_a$ comes from an error in the (v, b) parameters. If the target evolution operator corresponds to the $(v_0, b_0)$ values, the $U_a$ will include some small error and it will correspond to $(v_0+\delta v, b_0+\delta b)$. Assuming the errors $\delta v$ and $\delta b$ are small enough, their effect on U(T,-T) could be considered as lineal (by means of $\varepsilon$) in the phases and the probability p. The $U_a$ could be written as:

$$[U_a(T,-T)]_{00} = [U_a(T,-T)]_{11}^* =$$
$$= (1 - p + c_p\varepsilon) e^{i2(\alpha_{00}+c_0\varepsilon)} + pe^{i2(\alpha_{01}+c_1\varepsilon)}$$

$$[U_a(T,-T)]_{01} = [U_a(T,-T)]_{10} =$$
$$= -2i\sqrt{(1-p+c_p\varepsilon)(p+c_p\varepsilon)} \sin(\alpha_{00}+c_0\varepsilon - \alpha_{01} - c_1\varepsilon)$$

$$(20)$$

c's being real numbers.

By means of the $U_a(T,-T)$, the matrix $P = (U_a-U_t)^+(U_a-U_t)$ is calculated and its trace is developed in a $\varepsilon$-power series. For a small enough $\varepsilon$, the error propagation provides a gate error probability behaviour as $P_e \sim O(\varepsilon^2)$, showing the intrinsic fault-tolerance of the diabatic gates built. The coefficient of $\varepsilon^2$ is complicated and depends on p and the phases $\alpha_{ij}$, and will be shown for each particular gate.

In the following, the set of one-qubit Pauli gates {X = NOT, Z, Y = -iXZ}, T (called $\pi/8$ gate), H (Hadamard), CNOT and Toffoli gate will be considered. Through a possible Hamiltonian, the parameters (v, b) will be numerically estimated to implement the gates as well as the gate error probabilities.

3.2 Pauli gates

Working in the computation basis set {|0>, |1>}, the looked for evolution inside the black box for the NOT gate is |0> $\leftrightarrow$ |1>, meaning that if the ingoing state is |0>, the outgoing must be |1> and vice versa, both with certainty.

Suppose the initial (s = 1) state $|\chi_0>$ = |0> is introduced into the NOT-gate black box, after the evolution as a "collision" varying s = 1 $\rightarrow$ 0 $\rightarrow$ 1, a probability of one is required for the process $|\chi_0> \rightarrow |\chi_1>$ = |1> (the same for the reverse process). Two Hamiltonians are introduced: $H_0$ having the eigenstates {$|\chi_0>$ = |0>, $|\chi_1>$ = |1>} and H with an adiabatic basis {$|\phi_0>$, $|\phi_1>$}, fulfilling $|\phi_0(s=1)>$ = |0> and $|\phi_1(s=1)>$ = |1>. The |0> state evolves in the NOT-gate black box through the evolution operator (15), and the probability that the system exits through the $|\phi_1(s=1)>$ = |1> channel (or vice versa) is:

$$P_{|0\rangle \leftrightarrow |1\rangle} = 4(1-p)p \sin^2(\alpha_{00} - \alpha_{01}) \qquad (21)$$

To get the state |1> with certainty, a probability p ~ 0.5 and $\alpha_{00}$-$\alpha_{01}$ ~ m$\pi$/2 (m an integer), are needed. Notice the probability p as well as the phase difference depends on the collision parameters (v, b), producing a structure of the probability $P_{|0\rangle \leftrightarrow |1\rangle}$ similar to the well known Stückelberg oscillations [33] appearing in atomic collisions. The oscillation came from the two-way interference, producing the exit state as is shown in figure 2(a).

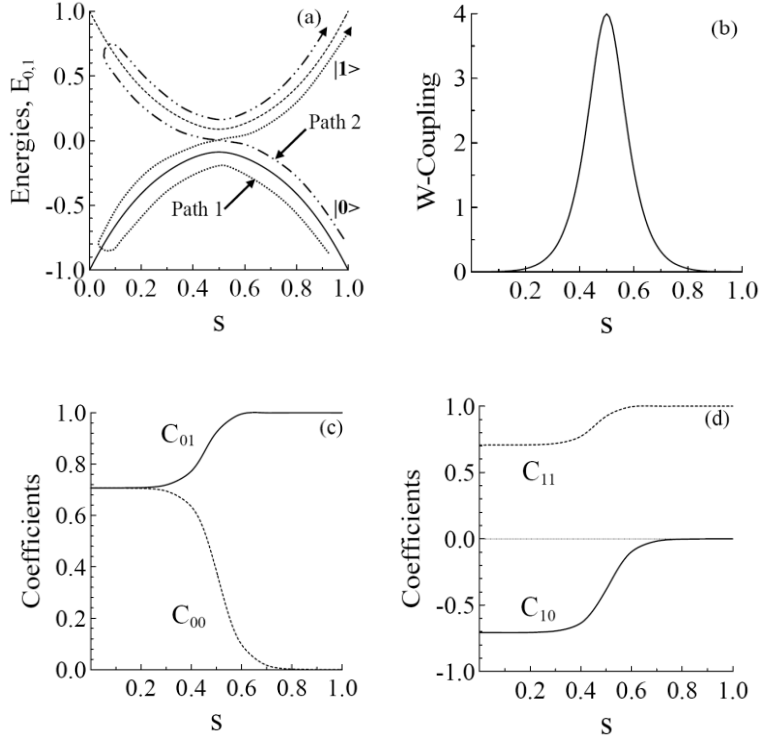

**Fig. 2** (a) Energy of adiabatic eigenstates $\{|\phi_0\rangle, |\phi_1\rangle\}$ of the total Hamiltonian describing a NOT-gate versus the parameter s. The initial states for s = 1 are shown on the right and the avoided-crossing will provide a radial coupling between them with a maximum at s ~ 0.5. The possible paths interfering to produce the $|\phi_1\rangle \equiv |1\rangle$ state are shown. (b) Radial coupling $W = \langle\phi_0|\partial/\partial s|\phi_1\rangle$. (c) and (d) Coefficients for the adiabatic states $|\phi_{i,k}\rangle = c_{(i\ or\ k),0}|0\rangle + c_{(i\ or\ k),1}|1\rangle$.

The next step must be the estimation of the (v, b) parameters to produce this gate. The impact parameter ($0 \leq b \leq s=1$) is related to how active the coupling W is in the collision, because it controls the minimum approach distance to the z = 0 point. The value b → 0 will be chosen to make the coupling completely active, and the most important parameter becomes the speed v. In the low-speed region, the adiabatic term in the system (14) will dominate the non-adiabatic one and, the evolution will be mainly adiabatic. In this region the adiabatic behaviour for quantum computation is recovered, whereas in the higher speed range, the non-adiabatic transitions will take place, thus raising the diabatic behaviour. In order to estimate the speed for which the maximum of $P_{|0\rangle \to |1\rangle}$ appears, the Massey parameter could be used. Instead of this, and supposing a first order solution for the initial condition $a_1(0) = \delta_{01}$, a more accurate estimation comes from the parameter:

$$\eta(v,b) = \left| \int_{z_{min}}^{z_{max}} W \frac{z}{s} \exp\left( \frac{i}{v} \int_{z_{min}}^{z} (E_1 - E_0)\, dz' \right) \right|^2, \qquad (22)$$

providing an approximation to $|b_1(z_{max})|^2$.

A possible Hamiltonian describing the "collision" could be as follows:

$$H = f(s) H_0 + g(s) H_W = s^4 \left(-|0\rangle\langle 0| + |1\rangle\langle 1|\right) - (1-s)^4 \left(|0\rangle\langle 1| + |1\rangle\langle 0|\right) \quad (23)$$

The functions $f(s) = s^4$ and $g(s) = -(1-s)^4$ are chosen only as an example to reach the objectives. By diagonalizing this Hamiltonian in the computation basis set $\{|0\rangle, |1\rangle\}$, the adiabatic basis $\{|\phi_0\rangle, |\phi_1\rangle\}$ is obtained, that will be used to expand the total state describing the dynamic:

$$\begin{aligned}|\phi_0(s)\rangle &= c_{00}(s)|0\rangle + c_{01}(s)|1\rangle \\ |\phi_1(s)\rangle &= c_{10}(s)|0\rangle + c_{11}(s)|1\rangle\end{aligned} \quad (24)$$

The coupling Hamiltonian introduced $H_W$, provides a radial coupling whose surface is $\pi/4$ and the adiabatic states comply with the asymptotic behaviour:

$$|+\rangle = \frac{1}{\sqrt{2}}(|0\rangle + |1\rangle) \xleftarrow{s \to 0} |\phi_0\rangle \xrightarrow{s \to 1} |0\rangle$$

$$-|-\rangle = \frac{1}{\sqrt{2}}(-|0\rangle + |1\rangle) \xleftarrow{s \to 0} |\phi_1\rangle \xrightarrow{s \to 1} |1\rangle \quad (25)$$

Figure 2 shows the results for the eigenvalues, "radial" coupling and coefficients $\{c_{ij}(s), i, j = 0, 1\}$. The ingoing state of the system is $|\phi_0(s=1)\rangle = |0\rangle$, and the objective is to reach an appropriate evolution (by adjusting the parameters v and b) to get a transition probability $P_{|0\rangle \to |\phi_1\rangle} \sim 1$ and $U(T,-T) \propto X$. Optimizing the parameters (v, b), the condition $\alpha_{00} - \alpha_{01} = m\,\pi/2$ (m an integer) is fulfilled and the gate $U(T,-T) = \pm i\,X$ can be reached, that is a NOT-gate except an unimportant global phase. Carrying out the numerical calculation, the optimized values are (v = 0.2547, b = 0). Figure 3 details the transition probability $P_{|0\rangle \to |\phi_1\rangle}$ versus v, for b = 0, showing a $P_{|0\rangle \to |\phi_1\rangle} = 0.99992$ for v ~ 0.2547. The required value of p ~ 0.5, reached for this speed, is shown in figure 4. In the same figure 3 the parameter η is included to appreciate its capability of estimating the speed (v ~ 0.249) for the maximum of the transition probability. For v < 0.1, typical Stückelberg oscillations appear as a consequence of a two-path interference producing the same outgoing state (see figure 2(a)). The envelope of the first oscillating part does not reach the value $P_{|0\rangle \to |\phi_1\rangle}$ ~ 1 because of the probability p < 0.5 for this speed.

Consequently, it is possible to get a high fidelity NOT-gate by means of this simple Hamiltonian involving only two coupled states. Because what is looked for is the transition between the adiabatic states (including the appropriate relative phases), the process could be called diabatic, and the gate, quantum diabatic NOT-gate.

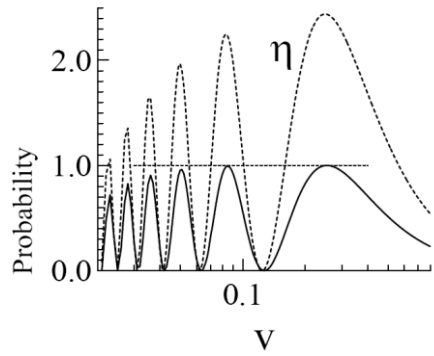

Fig. 3 Continuous line: Transition probability |0> → |1> ($P_{|0\rangle \to |\phi_1\rangle}$) in a two-state NOT-gate versus the speed v. The maximum value of the vertical axis is 1. The impact parameter b is taken as 0. Dashed line: η(v, b = 0), parameter providing an estimation of the maxima. In this case the vertical axis goes on to 2.5 arbitrary units.

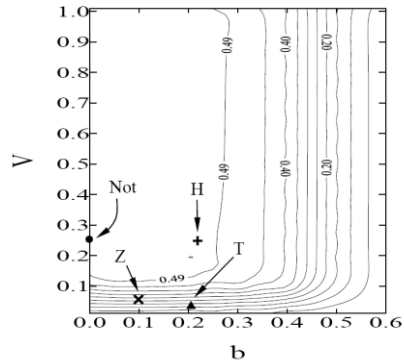

**Fig. 4** Probability p versus v and b. Several symbols represent the values for the optimized parameters (v, b) used for the quantum gates considered.

The gate error probability, calculated by means of tr{P} has the explicit structure:

$$P_e(\text{NOT}) = (2(c_0 - c_1)^2 + 8c_p^2)\varepsilon^2 + O(\varepsilon^3) \tag{26}$$

The $O(\varepsilon^2)$ behaviour for $P_e$ when v or b are changed has been checked. For instance, if the collision speed is changed around the optimum value $v_0$ in some $\varepsilon = \varepsilon_v$, the behaviour is $P_e \sim 40\ \varepsilon_v^2$. For ($v_0 = 0.2547$, $b = 0$), the gate error probability (Eq. (19)) given as the maximum eigenvalue of the P matrix, is $d_m(\text{NOT}) = 8\times10^{-5} \geq P_e$.

A similar study can be carried out with the remaining Pauli gates. For the Z gate, the chosen parameters ($v = 0.051$, $b = 0.1094$) fulfil the condition $\alpha_{00} = \pi/4$ and $\alpha_{01} = 5\pi/4$ and $U(T,-T) = i\,Z$ with $P_e(Z) \leq d_m(Z) = 5\times10^{-6}$. The general behaviour for the error gate probability is:

$$P_e(Z) = 8(c_p^2 - p(c_0^2 - c_1^2) + c_0^2)\varepsilon^2 + O(\varepsilon^3) \tag{27}$$

In this case, if the (v, b) parameters are correctly tuned; the gate does not depend on the probability p, so the condition $p \sim 0.5$ is not strictly needed, although the value of $P_e$ depends on p. The value of p for ($v = 0.051$, $b = 0.1094$) is shown in figure 4. The behaviour for $P_e$ when v is changed by keeping $b = 0.1094$ constant provides an error dependence $P_e \sim 4\times10^3\ \varepsilon_v^2$, $\varepsilon_v$ being the error for the speed.

The gate Y can be implemented by applying two successive gates according $Y = -i\,XZ$.

3.3 T gate

Another interesting one-qubit gate is the *T*-gate defined as:

$$T = \begin{pmatrix} 1 & 0 \\ 0 & e^{i\pi/4} \end{pmatrix} \tag{28}$$

This gate is also referred to as $\pi/8$ because, up to an unimportant $e^{i\pi/8}$ phase, it is equivalent to $\pi/8$-gate-phase, with:

$$T = e^{i\pi/8}\begin{pmatrix} e^{-i\pi/8} & 0 \\ 0 & e^{i\pi/8} \end{pmatrix} \tag{29}$$

To obtain the gate, the conditions $\alpha_{00} = 15\pi/16$ and $\alpha_{01} = -\pi/16$ must be fulfilled by optimizing (v, b). The general error gate probability is given by:

$$P_e(T) = (8c_p^2 + 4c_0^2 + 4c_1^2)\varepsilon^2 + O(\varepsilon^3) \qquad (30)$$

If the values of (v, b) are adequately tuned, the T-gate does not depend on the probability p, nor the value of $P_e$. The numerical calculation for (v = 0.0337, b = 0.2164) give the value $P_e \leq d_m(T) \sim 3\times10^{-6}$. The behaviour for $P_e$ when v is changed by keeping b = 0.2164 constant provides an error dependence $P_e \sim 2\times10^5 \varepsilon_v^2$, $\varepsilon_v$ being the error for the speed.

3.4 Hadamard gate

A Hadamard (in fact the gate iH) gate can be obtained by tuning the (v, b) parameters to fulfil the conditions $\alpha_{00} = 3\pi/8$ and $\alpha_{01} = -7\pi/8$ and p ~ 0.5 (shown in figure 4). The numerical calculation provides (v = 0.2249, b = 0.2677). For these values, the $P_e \leq d_m(H) \sim 3.5\times10^{-5}$. The behaviour for $P_e$ when v is changed by keeping b = 0.2677 constant gives an error dependence $P_e \sim 27 \varepsilon_v^2$, $\varepsilon_v$ being the error for the speed.

3.5 Control-Not and Toffoli gates

Surprisingly, the same kind of Hamiltonian can be used to implement several other (more complicated) gates as the CNOT gate. In this case the sought transitions are $|10\rangle \leftrightarrow |11\rangle$, whereas the states $|00\rangle$ and $|01\rangle$ should not be affected. A Hamiltonian representing this behaviour could be:

$$H = s^4 (-|10\rangle\langle10| + |11\rangle\langle11|) - (1-s)^4 (|10\rangle\langle11| + |11\rangle\langle10|) \qquad (31)$$

The same discussion is appropriate for the Toffoli-gate, in this case, the total Hamiltonian could be:

$$H = s^4 (-|110\rangle\langle110| + |111\rangle\langle111|) - (1-s)^4 (|110\rangle\langle111| + |111\rangle\langle110|) \qquad (32)$$

All the conclusions reached in the case of the NOT-gate are completely valid. Although in the case of Toffoli gate the Hamiltonian would involve three qubit interactions, there is no fundamental reason to not be considered.

In fact, to implement the gate only an eight state system is needed in which the behaviour was according (32).

**4. Is there a diabatic quantum computation model?**

Several one and two qubits gates have been represented as diabatic gates, identifying a simple theoretical Hamiltonian carrying out the work. In this situation, the concept of diabatic computation considered has been as a sequence of diabatic gates. Perhaps a further step could be made: could this diabatic gate model be extended to a full diabatic computation understood as a *single evolution* carried out by means of a suitable Hamiltonian? I guess the answer to this question is yes, because the general form of the evolution operator involving two-states (equation (15)) is completely determined providing three parameters (n(n+1)/2 with n=2), in our case p, $\alpha_{00}$ and $\alpha_{01}$. The only remaining (perhaps hard) work is to characterise the physical processes by implementing the required Hamiltonian. These Hamiltonians could involve non-local interactions of more than two or three qubits, and this could be seen as a heavy restriction nowadays, but is not forbidden at all by the laws of Quantum Mechanics. In addition, and as a consequence of the previous sections, this quantum diabatic computation should show some kind of *intrinsic* fault-tolerance against noise.

**5. Conclusions**

A new model for quantum computation is provided based on controlling the transition probabilities in a strong non-adiabatic evolution represented by an avoiding crossing. The model is introduced by means of the analogy with an evolution in an inelastic atomic collision and is described by means of a Hamiltonian having an implicit time-dependence through an s(t) parameter closely related to the internuclear atomic distance. A simple two-state model is enough to reproduce the quantum gates. The evolution operator is characterized by two parameters, the collision speed v and the impact parameter b, that must be adjusted in order to reproduce the desired quantum gate.

By using this method, several gates have been constructed, particularly the CNOT, Pauli, T, Hadamard and Toffoli, the sets being {CNOT, T, H} as well as {CNOT, H, Toffoli} universal. These gates have the advantage of an *intrinsic fault-tolerance* because the error gate probability behaves as $O(\varepsilon^2)$, $\varepsilon$ being some error with respect to the correct parameters ($v_0$, $b_0$). For each gate considered, the parameters (v, b) have been obtained in order to get an error gate probability smaller than $10^{-4}$, this value is considered as the threshold for doing fault-tolerant quantum computation. Evidently, if the experimental implementation is able to tune these parameters with higher precision, the gate error will reach smaller values.

As the present strategy looks for large transition probabilities, it could be called *diabatic quantum computation*, involving states going smoothly through an avoided-crossing. The conclusion is the possibility of doing universal diabatic fault-tolerant quantum computation; opening up a new possibility of implementing quantum gates and quantum algorithms experimentally.

Finally, and going a step further, a new quantum diabatic computation model is glimpsed, where a carefully chosen Hamiltonian could carry out a non-adiabatic transition between the initial and the sought final state. The problem would be to get a physical system evolving according this highly non-local Hamiltonian.

**Acknowledgment**

This work has been supported by the research project Quantum Information Technologies (QUITEMAD), P2009/ESP-1594 of Comunidad Autónoma de Madrid in Spain.